\documentclass[pra,twocolumn, floatfix, footinbib,superscriptaddress, longbibliography]{revtex4-2}
\usepackage[utf8x]{inputenc}
\usepackage[english]{babel} 
\usepackage[english]{babel} 
\usepackage[colorlinks=true,bookmarks=false,citecolor=blue,urlcolor=blue]{hyperref}

\usepackage{amsmath,amssymb,esint}
\usepackage{graphicx}
\usepackage{graphics}
\usepackage{verbatim}
\usepackage{bbold}
\usepackage[english]{babel}
\usepackage{bm}
\usepackage{natbib}
\usepackage[normalem]{ulem}
\usepackage{epstopdf}
\usepackage{multirow}
\usepackage{ dsfont }
\usepackage{tikz}
\DeclareMathAlphabet\mathbfcal{OMS}{cmsy}{b}{n}

 \usepackage{booktabs}
 \usepackage{multirow}
\usepackage{ragged2e}

\usepackage{float}

\usepackage{amsmath}
\usepackage[colorinlistoftodos]{todonotes}
\usepackage{color}
\graphicspath{{fig/}}
\usepackage{bm}
\usepackage{braket}
\newcommand{\ve}[1]{\mathbf{#1}}
\renewcommand{\t}[1]{\text{#1}}
\renewcommand{\Re}{\operatorname{Re}}
\renewcommand{\Im}{\operatorname{Im}}

\newcommand{\te}[1]{\widehat{{#1}}}

\newcommand{\red}[1]{{\color{red}#1}}

\let\vec=\mathbf

\begin{document}

\title{
Directional emission of down-converted photons from a dielectric nano-resonator}

\newcommand{\affilITMO}{ITMO University, Birzhevaya liniya 14, 199034 St.-Petersburg, Russia}

\newcommand{\affilUTS}{School of Mathematical and Physical Sciences, University of Technology Sydney, Sydney NSW 2007, Australia}

\author{Anna Nikolaeva}
\email{anna.nikolaeva@metalab.ifmo.ru}
\affiliation{\affilITMO}
\author{Kristina Frizyuk}
\affiliation{\affilITMO}
\author{Nikita Olekhno}
\affiliation{\affilITMO}
\author{Alexander Solntsev}
\affiliation{\affilUTS}
\author{Mihail Petrov}
\affiliation{\affilITMO}
\date{\today}

\begin{abstract}
Creation of correlated photon pairs is one of the key topics in contemporary quantum optics. Here, we theoretically describe the generation of photon pairs in the process of spontaneous parametric down-conversion in a resonant spherical nanoparticle made of a dielectric material with bulk $\hat \chi^{(2)}$ nonlinearity. We pick the nanoparticle size that satisfies the condition of resonant eigenmodes described by Mie theory. We reveal that highly directional photon-pair generation can be observed utilising the nonlinear Kerker-type effect, and that this regime provides useful polarisation correlations.\end{abstract}


\maketitle

\section{\label{sec:intro}Introduction}
Spontaneous parametric down-conversion (SPDC) is arguably the most widely used process for the generation of non-classical light~\cite{1970JETP}. On one hand, compared to quantum dots~\cite{Muller:2014-224:NPHOT, Versteegh:2014-5298:NCOM} and atomic defects~\cite{RN366}, SPDC does not need cryogenic cooling to produce pure and indistinguishable heralded single photons~\cite{Massaro_2019} and correlated pairs of photons~\cite{Solntsev:2017-19:RPH, PhysRevA.92.033815} at a high rate. It is a major advantage, since optical helium cryostats are incredibly bulky and expensive, thus preventing the proliferation of non-classical light sources that require them. On the other hand, until recently, one of the disadvantages of SPDC compared to its more compact photon source counterparts was a requirement for a significant volume of nonlinear material. The SPDC source footprint was limited to tens or hundreds of microns~\cite{Guo2017}. Another major disadvantage was the necessity to maintain stringent phase matching, which often required precise temperature control~\cite{Setzpfandt2016}. 

With the certain progress \cite{DinparastiSaleh2018,Okoth2019,Li2020} in overcoming these limitations in subwavelength nanophotonic systems along with the advent of Mie-resonant nano-photonics based on nonlinear dielectric nanoparticles~\cite{Camacho-Morales:2016-7191:NANL, Won2019}, both of these disadvantages are now being lifted~\cite{1903.06956}. The Mie resonant nanostructures have already been suggested to enhance spontaneous photon emission process such as Raman scattering \cite{Frizyuk2018,Dmitriev2016a}, photoluminescence emission \cite{Zambrana-puyalto2015,Rocco2020} as well as SPDC process~\cite{Poddubny2018}. One of the important remaining roadblocks in this space is a directionality control of SPDC emission. At the same time, Mie-photonic structures provide unique tools for controlling radiation pattern due to interplay between the  magnetic and electric resonant modes. An useful example of electric and magnetic modes interference is a so-called Kerker effect, which enables strong forward scattering of light dielectric particles~\cite{Staude2013, Liu_Kivshar_2018, Tzarouchis_Sihvola_2018, Kerker_Wang_Giles_1983}. This feature has been also utilized for the manipulation of reflection, transmission, diffraction, and absorption  for metalattices and metasurfaces~\cite{Liu_Kivshar_2018, Babicheva2017,Moitra2015,Decker2015}. Moreover, the proper engineering of the resonant modes allows for tailoring of the far-field pattern of the generated signal~\cite{Saerens2020}. The quantum analogue of Kerker-effect has more complexity, since correlations between signal and idler photons generated via SPDC have to be taken into account~\cite{1970JETP}. This complexity however also offers significant flexibility.

In this work, we develop a detailed theoretical approach for SPDC in a Mie-resonant dielectric nano-resonator utilising the Kerker-type effect. We focus on nonlinear two-photon decay in a spherical nanoparticle made of a material with bulk quadratic optical susceptibility $\hat\chi^{(2)}$ in classical Mie geometry. We show that highly directional SPDC generation can be achieved based on the interplay between electric and magnetic dipole emission, featuring significant flexibility depending on the polarization orientations. We also demonstrate that useful signal-idler polarization correlations can be generated in this system.

\begin{figure}[t]
{\includegraphics[width=0.6\linewidth]{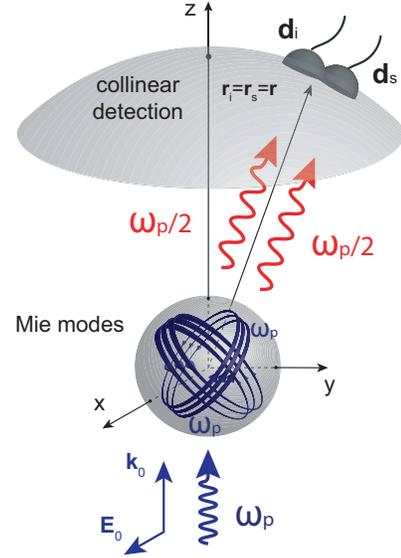}}
\caption{Schematic of SPDC process in collinear decay geometry: signal and idler photons are being detected at the point in the far-field $\ve r_i=\ve r_s=\ve r$.}
\label{fig:1_SPDC}
\end{figure}

The manuscript is organized as follows: in Section~\ref{sec:theory} we firstly give a brief overview of classical Kerker-effect and directional elastic scattering (linear scattering at the frequency of the incident field) of photons by a spherical nanoparticle. Secondly, we propose an approach to analysis of SPDC process based on Green's function method and its multipolar representation for constructing two-photon amplitude. In Section~\ref{sec:GaAs_np}, we apply our approach for analyzing SPDC emission from GaAs nanoparticles and show that in case of Wurzite crystalline structure one can expect pronounced forward/backward scattering of entangled photons within the conditions closely related to standard Kerker-effect. We also discuss the alternative crystalline systems, which can provide strongly directional emission of entangled photons. And finally, we consider polarization correlations in the emitted photons.

\section{Theoretical  framework}
\label{sec:theory}
\subsection{Mie resonances and Kerker-type elastic scattering}
\label{sec:theory_Kerker}
We start with our consideration with reviewing the conditions for preferential forward elastic scattering of light by a nanoparticle and  interpretation of the elastic Kerker effect in  terms of the Mie modes. According to the Mie theory \cite{Bohren1998}, expression for the scattered electric field $\vec{E}_s$  and electric field inside nanoparticle  $\vec{E}_p$ can be expressed as
\begin{align}
\vec{E}_{s}(k_1, \vec{r})=\sum\limits_{n=1}^{\infty} E_{n}(ia_n\vec{N}^{(1)}_{e1n}(k_1, \vec{r})-b_n\vec{M}^{(1)}_{o1n}(k_1, \vec{r})),
\label{eq:E_s}
\end{align}
\begin{align}
\vec{E}_p(k_2,\vec{r})=\sum\limits_{n=1}^{\infty}E_n(c_n\vec{M}_{o1n}(k_2, \vec{r})-id_n\vec{N}_{e1n}(k_2, \vec{r}))
\label{eq:E_p}
\end{align}
where $E_{n}= i^n E_{0}{(2n+1)}/{(n(n+1))}$ with $E_0$ {and $k_1(\omega)=\sqrt{\varepsilon_1}\omega/c$  being the amplitude and the wave number of the incident plane wave, $k_2(\omega)=\sqrt{\varepsilon_2(\omega)}\omega/c$ is the wave number inside the nanoparticle, $\vec{M}^{}_{^{e}_{o}mn}$ and $\vec{N}^{}_{^{e}_{o}mn}$ are respectively the magnetic and the electric  vectorial spherical harmonics (VSH)(see Appendix~\ref{app:VSH}). Here we append the superscript (1) to vector spherical harmonics for which the radial dependence of the generating functions is specified by spherical Hankel function $h_n^{(1)}(x)$ instead of  spherical Bessel function $j_n^{}(x)$. The incident plane wave propagates along  $z$-direction with the polarization along  $x$-axis. We limit our consideration by dipole modes only, $n=1$, so we can simplify notations for the  indexes of the VSHs: $e11 =x; \ \ o11=y; \ \ e01=z$}. The Kerker effect can  interpreted in terms of the field structure and symmetry of these harmonics \cite{Fu2013}. {In the simplest case}, directional light scattering arises due to the crossed magnetic and electric dipole which contribute to the field with the different phases. The scattering intensity is described by Poynting vector $\ve S $, which in the dipolar approximation provides:

\begin{gather}
|\ve S(\ve r)|\simeq |\ve E_s(\ve r)|^2=|E_1|^2\left[|a_1|^2|\vec{N}^{(1)}_{x}(\ve r)|^2+|b_1|^2|\vec{M}^{(1)}_{y}(\ve r)|^2\right.\nonumber \\
\left.+2\Re (ia_1^*b_1\vec{N}^{(1)*}_{x}(\ve r)\vec{M}^{(1)}_{y} (\ve r))\right]\nonumber \\
=|E_1|^2\left[|a_1|^2|\vec{N}^{(1)}_{x}(\ve r)|^2+|b_1|^2|\vec{M}^{(1)}_{y} (\ve r)|^2\right.\nonumber \\
\left. -2|a_1^*b_1||\vec{N}^{(1)*}_{x}(\ve r)\vec{M}^{(1)}_{y} (\ve r)|\cos(\varphi_{b_1}-\varphi_{a_1})\sin\left(\varphi_{\vec{M}_y}-\varphi_{\vec{N}_x}\right)\right.\nonumber\\
\left.-2|a_1^*b_1||\ve {N}^{(1)*}_{x}(\ve r)\ve {M}^{(1)}_{y}(\ve r)|\cos(\varphi_{\ve M_y}-\varphi_{\ve N_x})\sin(\varphi_{b_1}-\varphi_{a_1})\right]
\label{eq:Direct_elastic}
\end{gather}
Here $\varphi_{a_1}, \varphi_{b_1}, \varphi_{\ve M_y}$, and $ \varphi_{\ve N_x}$ define correspondingly the phase of the complex valued $a_1$, $b_1$ coefficients and  $\ve{N}^{(1)}_{x}$, $\ve{M}^{(1)}_{y}$ vector harmonics. One can see that the interference term depends on both relative phases of: i) electric and magnetic dipole amplitudes $a_1$ and $b_1$ and ii) corresponding vector harmonics $\vec{M}^{(1)}_{y}$ and $\vec{N}^{(1)}_{x}$.  The latter contribution in the far-field domain $kr\gg 1 $ provides that{ $\varphi_{\vec{M}_y}-\varphi_{\vec{N}_x}=-{\pi}/{2}$ for $\theta=0$, while  $\varphi_{\vec{M}_y}-\varphi_{\vec{N}_x}={\pi}/{2}$ for $\theta=\pi$}, where $\theta$ is the polar angle in spherical coordinate system. Moreover, since the amplitudes of  $\ve M^{(1)}_{y}$ and $\ve N^{(1)}_{x}$ harmonics at points $\theta =0, \pi$ are equal due to the symmetry of magnetic and electric dipole emission  and using {$\left|\vec{N}^{(1)}_{x}(\theta=0, \pi)\right|^2=\left|\vec{M}^{(1)}_{y}(\theta=0, \pi)\right|^2=\left|\vec{N}^{(1)*}_{x}(\theta=0, \pi)\vec{M}^{(1)}_{y}(\theta=0, \pi)\right|$}, we come to a simple expression:
\begin{gather}
S(\theta=0,\pi)\sim|E_1|^2\left[|a_1|^2+|b_1|^2
\pm2|a^*_1b_1|\cos\left(\varphi_{b_1}-\varphi_{a_1}\right)\right].
\label{eq:Linear_Kerker}
\end{gather}

Basing on that, one immediately comes to a well-known Kerker condition for strong forward scattering $|a_1|=|b_1|$ and $\varphi_{a_1}=\varphi_{b_1}$, or for strong backward scattering (second Kerker's condition) $|a_1|=|b_1|$ and {$\varphi_{b_1}-\varphi_{a_1}=\pi$}. These simple results lie in the basis of directionality engineering with resonant dielectric structures. 

\subsection{Nonlinear generation of entangled photons }

In the course of SPDC nonlinear process, a photon with frequency $\omega_p$ is absorbed and two photons with frequencies $\omega_i$ (idler photon) and $\omega_s$ (signal photon) are generated \cite{1970JETP} in a way that the energy should be conserved   $\hbar\omega_p = \hbar\omega_i+\hbar\omega_s$. In the bulk medium, where the photons can be described with a plane waves, the momentum conservation law (phase matching condition) should also be fulfilled $\ve k_p=\ve k_i + \ve k_s$. However, in the subwavelength structures the latter condition can be violated~\cite{Okoth2019}, and the photon generation should be described in spherical multipoles basis which is more natural for such geometry. In order to apply such technique, one can describe SPDC process with a two-photon amplitude  $T_{is}(\vec{r}_i,\omega_i,\ve d_i,\vec{r}_s,\omega_s,\ve d_s)$, expressing the probability of simultaneous detection of idler and signal photons having correspondingly frequencies $\omega_i$ and $\omega_s$ at coordinates $\ve r_i$ and $\ve r_s$ with polarizations defined by the dipole moments of the detectors $\ve d_i$ and $\ve d_s$. According to the approach developed in Ref.  \cite{Poddubny2016}, the two-photon amplitude has the following form: 

\begin{align}
\label{eq:T_general}
& T_{is}(\vec{r}_i,\omega_i,\vec{d}_i;\vec{r}_s,\omega_s,\vec{d}_s)=\nonumber\\
 & = \int\limits_V \ve d_i^*{\vec{\hat{G}}(\vec{r}_i,\vec{r}_0,\omega_i)} {\hat{\Gamma}(\vec{r}_0)}{\vec{\hat{G}}(\vec{r}_0,\vec{r}_s,\omega_s)} {\ve d_s^*}d^3r_0,
\end{align}
where $\hat{\ve G}(\vec{r},\vec{r}_0,\omega)$ is the dyadic Green's function of the generating system, ${\Gamma}_{\alpha\beta}(\vec{r}_0)=\chi_{\alpha\beta\gamma}^{(2)}{E}^{\gamma}_p(\vec{r}_0)$ is the generation tensor, $\chi_{\alpha\beta\gamma}^{(2)}$ is the second-order nonlinear susceptibility tensor\cite{Boyd2003}, and $\vec{E}_p(\omega_p, \vec{r}_0)$ is the pump field which causes the nonlinear generation. Then, the probability of simultaneous detection of idler and signal photons of particular polarizations provided by the direction of $\ve d_i$ and $\ve d_s$ will be as follows
\begin{equation}
w_{\ve d_i,\ve d_s}=\dfrac{2\pi}{\hbar}\delta(\hbar\omega_i+\hbar\omega_s-\hbar\omega_{p})|T_{is}|^2.
\label{eq:W_pol}
\end{equation}

In the following, we will focus mainly on the unpolarized detection, when the amplitude can be obtained by the direct summation of Eq.~\eqref{eq:W_pol} over all possible polarizations of detectors:
$w^{\t{unpol}}=\sum_{\ve d_i,\ve d_s}w_{\ve d_i,\ve d_s}.$ We will also discuss the polarized entanglement in Sec.\ref{sec:Polarization} for particular orientations of the detectors.


\begin{figure}[t]
{\includegraphics[width=\linewidth]{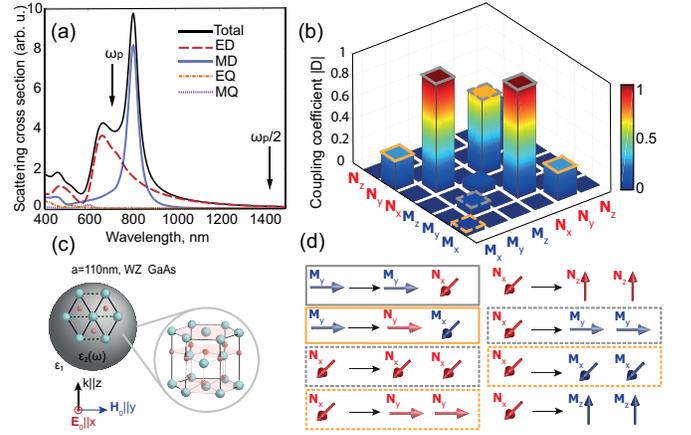}}
\caption{{\red{}(a) Elastic scattering cross-section, including different multipoles (solid black - total, dashed red- electric dipole ED, solid blue -magnetic dipole MD, dashed-dot yellow - electric quadrupole EQ, dotted purple - magnetic quadrupole MQ) depending on fundamental wavelength $\lambda_p$. (b) $D$-coefficients normalized to the maximum for all possible dipole decays at a wavelength $\lambda_p=720$ nm. (c) Geometry of the considered problem. Spherical particle of wurtzite GaAs with a radius $a=110$ nm, the incident plane wave propagates along the z-axis, the electric field oscillates along the x-axis, the figure also shows the orientation of the crystal lattice relative to the incident pane wave. (d) Possible decay channels in the considered geometry, gray or yellow needed to obtain directivity: solid - decay to the crossed dipoles, dashed - decay to the same dipoles.}}
\label{fig:2_GaAs}
\end{figure}

Next, we will provide the grounds for multipolar photon entanglement of the generated photons. In the case of nonlinear decay in bulk medium, there is particular spatial and polarization correlation between the free-space photons described by plane waves. In the case of subwavelength scale nonlinear source, for which plane wave description is substituted by multipolar harmonics, there  appears particular entanglement and correlations between the multipolar content of the generated photons. We will introduce additional notation to make the description of the considered process more compact and elegant. First of all, we will refer to vector spherical harmonics as $\ve{W}_{p_ip_rmn}(k,\vec{r})$ notation, where $m$ and $n$ are azimuthal and main quantum numbers, ${p_i}$ and $p_r$ are the inversion and reflection in the $y=0$ plane parities indexes \cite{Frizyuk2019, Gladyshev_Frizyuk_Bogdanov_2020}. We will define the set of four labels as $\ve J=\{p_i,p_r,m,n\}$. {Here, ${p_i}$ implicitly sets the polarization of the multipole, namely, if it is magnetic (M) or electric (N).  } Thus, the nonlinear decay process can be understood as a set of mulitpolar decay channels each characterized by J-vectors  $\ve J_{\t{pump}}\rightarrow \ve J_{\t{idler}}, \ve J_{\t{signal}} $.  In order to analyze it, we will expand the two-photon amplitude into particular multipolar channels. This can be done straight forwardly by expanding the Green's function (see Appendix~\ref{app:VSH}) and plugging in Mie-solution (\ref{eq:E_p}) already written in VSH basis resulting in the expression:

\begin{equation}
\begin{aligned}
&T_{is}(\vec{r}_i,\omega_i,\vec{d}_i;\vec{r}_s,\omega_s,\vec{d}_s)=\sum\limits_{\ve J_p, \ve J_i, \ve J_s} \widetilde T_{\ve J_p \rightarrow \ve J_i, \ve J_s}\times \\&
\times D_{\ve J_p \rightarrow \ve J_i, \ve J_s} \left(\ve d_i^*\cdot \ve W^{(1)}_{\ve J_i}(k_{i1},\vec{r_i})\right)\left(\ve W^{(1)}_{\ve J_s}(k_{s1},\vec{r_s})\cdot \ve d^*_s\right),
\end{aligned}
\label{eq:T_harmonics}
\end{equation}
where index 1 mean outside nanoparticle and, hence, we use $k_{i1(s1)}=k_1(\omega_{i(s)})$, index 2 mean inside nanoparticle, hence, we use  $k_{i2(s2)}=k_2(\omega_{i(s)})$ and

\begin{gather}
\widetilde T_{\ve J_p \rightarrow \ve J_i, \ve J_s}=-{\Big(\dfrac{\omega_i\omega_s}{c^2}\Big)^2} {k_{i2}k_{s2}}A_{\ve J_i}A_{\ve J_s}B_{\ve J_p},\\
A_{\ve J}=(2-\delta_0)\dfrac{2n+1}{n(n+1)}\dfrac{(n-m)!}{(n+m)!}\cdot\begin{cases}
a_n^{(2)}\text{,~if}~{p_i \rightarrow M},  \\b_n^{(2)}\text{,~if}~{p_i \rightarrow N}, 
\end{cases} \\
B_{\ve J	}=E_n\cdot\begin{cases}
c_n\text{,~if}~{p_i \rightarrow M},  \\-id_n\text{,~if}~{p_i \rightarrow N}. 
\end{cases} 
\end{gather}
Coefficient $D_{\ve J_p \rightarrow \ve J_i, \ve J_s}$ denotes  the intensity of the particular decay channels, and contains a spherical harmonic from the pumping field expansion and two harmonics from the  decay field included in dyadic Green's function:

\begin{align}
&D_{\ve J_p \rightarrow \ve J_i, \ve J_s}= \sum_{\alpha,\beta, \gamma}\chi_{\alpha\beta\gamma}^{(2)} \times \nonumber\\ 
&\int\limits_V{ W}_{ \ve J_p, \gamma}(k_2,\vec{r_0})  {W}_{ \ve J_i,\alpha}(k_{i2},\vec{r_0}) {W}_{  \ve J_s,\beta}(k_{s2},\vec{r_0})  d^3r_0, \label{eq:D-coeff} 
\end{align}
where ${W}_{  \ve J,\alpha}(k,\vec{r})$ is the Cartesian projection of the vector spherical harmonic $\ve W_{\ve J}(k,\vec{r})$ on the $\alpha$-axis, $\alpha,\beta,\gamma=x,y,z$. The D-coefficients \eqref{eq:D-coeff} are the overlapping integrals and determine the amplitude of particular decay channel. The non-zero values of D-coefficients provide allowed transitions and define the so-called selection rules of second order nonlinear process, which detailed symmetry based analysis is provided elsewhere  \cite{Frizyuk2019,Frizyuk_2019} for the inverse process of sum frequency generation. Below, we will discuss in details, what particular decay channels can contribute into the directional emission of entangled photons. 

We will restrict our consideration to the case of  simultaneous  detection of the correlated photons in preferable direction. We firstly  are interested in the total coincidence rate $w^{\t{unpol}}$ which accounts on detection of photon disregarding their polarization:

\begin{widetext}
  \begin{gather}
w_{is}^{\t{unpol}}(\vec{r}_i,\omega_i;\vec{r}_s,\omega_s)=\dfrac{2\pi}{\hbar} \sum\limits_{\ve d_i, \ve d_s} \left|T_{is}(\vec{r}_i,\omega_i,\vec{d}_i;\vec{r}_s,\omega_s,\vec{d}_s)\right|^2= \dfrac{2\pi}{\hbar}\sum\limits_{\ve d_i, \ve d_s}
 \sum\limits_{\substack{\ve J_p, \ve J_i, \ve J_s\\ \ve J'_p, \ve J'_i, \ve J'_s }}  \underbrace{\widetilde T_{\ve J_p \rightarrow \ve J_i, \ve J_s}\cdot D_{\ve J_p \rightarrow \ve J_i, \ve J_s}\cdot\widetilde T^*_{\ve J'_p \rightarrow \ve J'_i, \ve J'_s}\cdot  D^*_{\ve J'_p \rightarrow \ve J'_i, \ve J'_s}}_{C^{\ve J_p \rightarrow \ve J_i, \ve J_s}_{ \ve J'_p \rightarrow \ve J'_i, \ve J'_s}}\times \nonumber\\ 
\times \left(\ve d_i^*\cdot \ve W^{(1)}_{\ve J_i}(k_{i1},\vec{r_i})\right)\left(\ve W^{(1)}_{\ve J_s}(k_{s1},\vec{r_s})\cdot \ve d_s^*\right)\times
\left(\ve d_i\cdot \ve W^{*(1)}_{\ve J_i'}(k_{i1},\vec{r_i})\right)\left(\ve W^{*(1)}_{\ve J_s'}(k_{s1},\vec{r_s})\cdot \ve d_s\right) \nonumber\\
=
 \dfrac{2\pi}{\hbar}\sum\limits_{\substack{\ve J_p, \ve J_i, \ve J_s\\\ve J'_p, \ve J'_i, \ve J'_s}} C^{\ve J_p \rightarrow \ve J_i, \ve J_s}_{ \ve J'_p \rightarrow \ve J'_i, \ve J'_s} \left( \ve W^{(1)}_{\ve J_i}(k_{i1},\vec{r_i})\cdot \ve W^{*(1)}_{\ve J_i'}(k_{i1},\vec{r_i}) \right)\left(\ve W^{(1)}_{\ve J_s}(k_{s1},\vec{r_s})\cdot \ve W^{*(1)}_{\ve J_s'}(k_{s1},\vec{r_s}) \right)
\label{eq:Direct_nonl}
\end{gather}
\end{widetext}

The final form containing the scalar products of multipole VSH was obtained due to summation over the all polarizations of photons and more detailed derivation of it the expression for $C$-coefficients is provided in Appendix \ref{app:Derivation_T}. 
The obtained expression can be considered as an analogue of expression  \eqref{eq:Direct_elastic} but for nonlinear generation of the correlated photons. Each component in the sum consists of the multiplication of two idler photons ($\ve J_i$ and $\ve J_i'$) and two signal photons ($\ve J_s$ and $\ve J_s'$)  as all of them can potentially interfere due to coherence of the SPDC process.

\begin{figure}[h]
\center{\includegraphics[width=1\linewidth]{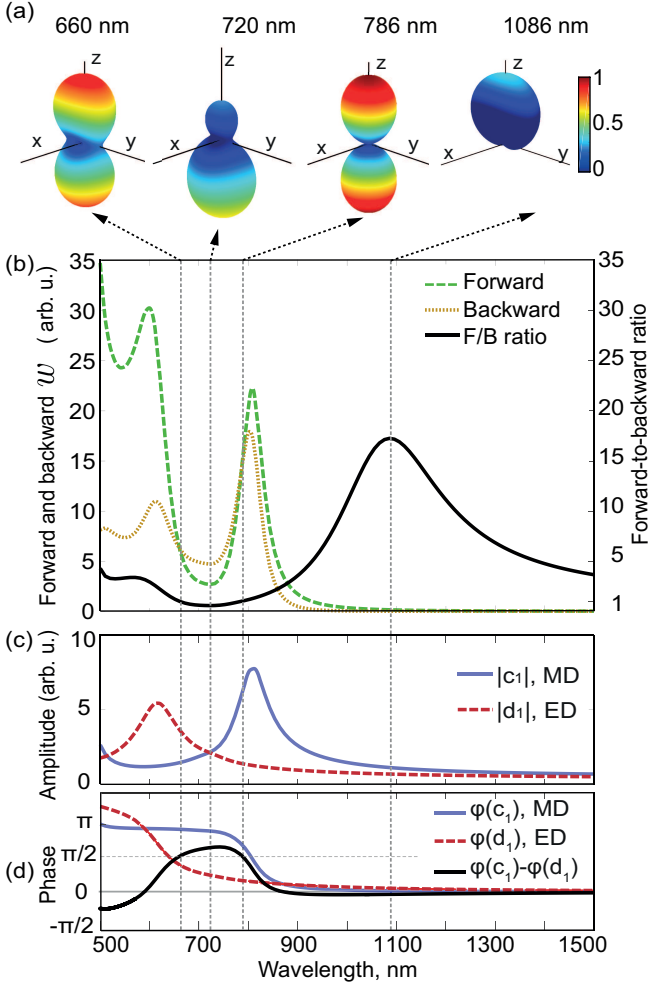}}
\caption{(a) The far-field patterns of collinear two-photon generation for different wavelengths $\lambda_p=$ 660~nm, 720~nm, 786~nm, 1086~nm. (b) Forward (dashed green) and backward (dotted yellow) $w_{unpol}$ and their ratio $w^{\t{unpol}}_{\theta=0}/w^{\t{unpol}}_{\theta=\pi}$ (solid black) depending on pump wavelength $\lambda_p$. (c) Amplitudes of coefficients $|c_1|$ (solid blue) and $|d_1|$ (dashed red) in decomposition of pump field inside nanoparticle $\ve E_p$. (d) Phases of this coefficients $\varphi_{c_1}$ (solid blue), $\varphi_{d_1}$ (dashed red) and their phase difference $\varphi_{c_1}-\varphi_{d_1}$ (solid black).}
\label{fig:3_direct_pattern}
\end{figure}
\section{Directional emission from GaAs nanoparticle}
\label{sec:GaAs_np}
We apply the theoretical background overview in the previous section to analyzing the nonlinear generation of correlated photons from a  nanoparticle made of a material with strong bulk nonlinearity such as GaAs. The III-V semiconductor materials have already been widely used for second-harmonic generation in a number of experiments \cite{Koshelev_Kruk_Melik-Gaykazyan_Choi_Bogdanov_Park_Kivshar_2020, doi:10.1021/acs.nanolett.9b01112, Saerens2020, doi:10.1021/acs.nanolett.8b00830, C8NR08034H, doi:10.1021/acs.nanolett.6b03525, doi:10.1021/acsnano.9b07117, doi:10.1021/acs.nanolett.7b01488,Melik-Gaykazyan2019, Roleoft, Carletti_2017, Gili:16, Ghirardini:17, doi:10.1021/acsphotonics.8b00810,Saerens2020} showing particular enhancement of nonlinear signal generation owing to high-refractive index and pronounced Mie  resonances in the visible and infrared region. Indeed, the elastic scattering spectrum of a spherical particle of $a=110$ nm radius  shown in Fig.\ref{fig:2_GaAs}(a) demonstrates electric (ED) and magnetic (MD) dipole resonances. Their interference results in directional (elastic) scattering  in accordance to Kerker effect overviewed in Sec.\ref{sec:theory_Kerker}. In further, we apply several limitations, which on one hand will simplify the model, and on the other will uncover all the necessary aspects  of the directional emission of correlated photons: i) among the variety of the detection geometries, we will specify a {\it collinear } detection, when the idle and signal detectors are located in the same point $\ve r_i=\ve r_s$ (see Fig.\ref{fig:1_SPDC})  ii) we focus on the  resonant pumping scheme, when the the frequency of the exciting field falls within the range  electric or magnetic dipole resonances; iii) we consider degenerate process assuming $\omega_i=\omega_s=\omega_p/2$, thus the generated photons will be far from the Mie resonance in the longwavelength region; iv) we neglect the birefringence of GaAs material, which does not provide strong contribution to second-order nonlinear process \cite{Frizyuk2019}; v) for the sake of definiteness, we will be interested in the strong directivity in forward/backward directions  of collinear emission of correlated photons, i.e. along $z$-axis. The latter provides  a more clear and simple narrative at this stage, however will be generalized in the following. 
\begin{figure}[t]
\center{\includegraphics[width=1\linewidth]{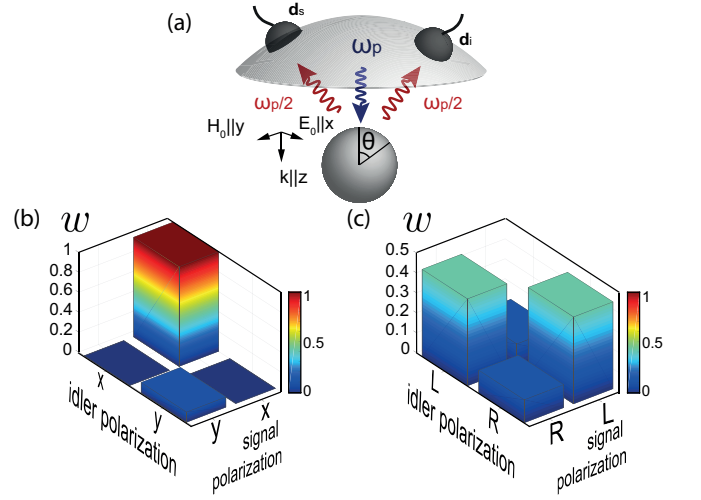}}\caption{a) Configuration of detection in the area limited by the maximum angle $\theta=60^{\circ}$ relative to the backward direction, radius of the sphere and all other parameters are similar as at \ref{fig:2_GaAs} (b, c) polarization correlations at $\lambda_p=720$nm in linear basis (b) and in circular basis (c)}
\label{fig:Pol_corr}
\end{figure}

The nonlinear response of GaAs is defined predominantly by the structure of its nonlinear tensor $\te \chi^{(2)}$. For a Wurzite crystalline structure the only non-zero elements of the tensor are  $\chi_{zzz}=  115$~pm/V,  $\chi_{zxx} = \chi_{zyy} =21$~pm/V, $\chi_{xxz} = \chi_{yyz} =42 $~pm/V \cite{Timofeeva2016}. The particular form of the tensor dictates the selection rules of the SPDC process through the non-zero $D$-coefficients. For the proposed form of the nonlinear tensor, the non-zero dipole decay channels are summarized in Fig.\ref{fig:2_GaAs}(b,d) under the assumption that the pumping field generates the $x$-oriented ED and $y$-oriented MD dipoles. These  channels define the non-zero components in the coincidence rate \eqref{eq:Direct_nonl}, while the scalar products of the vector functions corresponded to these modes, define the far-field patter of $w^{\t{unpolar}}$.


\begin{figure*}[t]
\center{\includegraphics[width=1\linewidth]{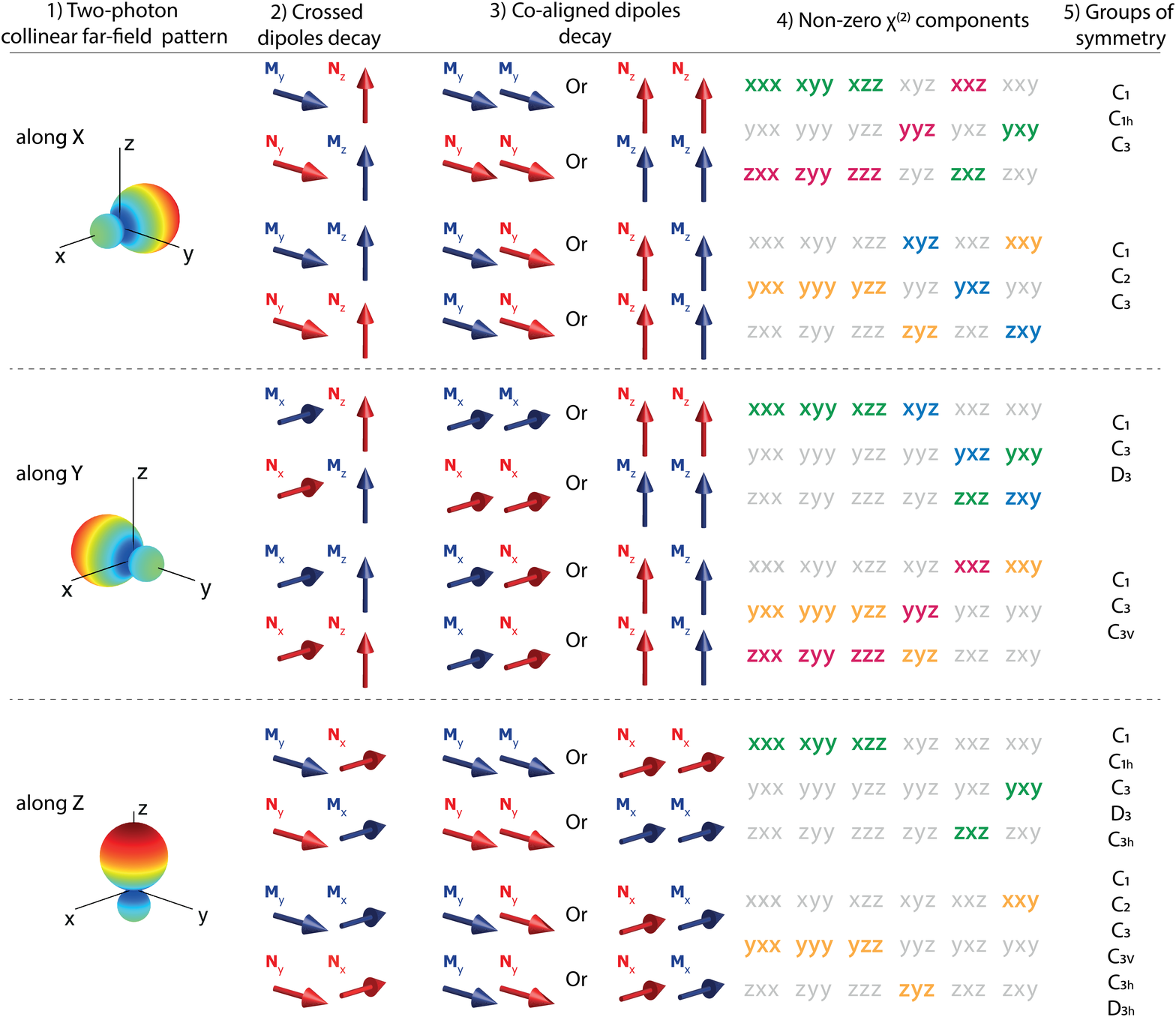}}
\caption{The structure of the nonlinear tensor providing the directivity of collinear two-photon emission in the specified direction. The columns: 1) far-field pattern of two-photon collinear detection along one  $x, y$, or $ z$ axes; 2) decay channels into crossed dipoles; 3)decay into co-directionall dipoles; 4) nonzero components of the second-order nonlinear susceptibility tensor $\hat\chi^{(2)}$; 5) crystal symmetry groups required to observe directivity.}
\label{fig:Table_chi2}
\end{figure*}

\subsection{Forward/backward directivity} As mentioned above, we are interested in $z$-directivity of collinear detection, i.e. when two correlated photons are generated either in forward or backward direction with respect to the direction of the exciting pump. In this formulation, this will provide us with a two-photon  analogue of the Kerker effect. Thus, careful analysis of all different contributions of dipole VSH scalar products in Eq.~\eqref{eq:Direct_nonl} schematically summarized in Appendix, Fig.~\ref{fig:app_scalar}, showing that only two components give contribution to the forward and backward directivity, which are scalar products $\ve M^{(1)}_y \cdot \ve N^{(1)}_x $ and $\ve M^{(1)}_x \cdot \ve N^{(1)}_y$. Thus, we derive an expression for the asymmetry of the detection rate $\Delta w^{\t{unpol}}=w^{\t{unpol}}_{\theta=0}-w^{\t{unpol}}_{\theta=\pi}$ in the proposed configuration. By taking into account that the decay of the photons is out of the resonance with the  Mie modes (see  Fig.\ref{fig:2_GaAs}), it can be assumed that the phase difference between the ED and MD coefficients at  the decay frequency $\omega_p/2$ is  zero   $\Delta\varphi^{decay}= \varphi_{a_1}-\varphi_{b_1}\approx 0$, as shown in Fig.~\ref{fig:1_SPDC} (c). Then the expression of the asymmetry can be simplified  as follows (See details in Appendix \ref{app:Derivation_W}) 
  
\begin{align}
& \Delta w^{\t{unpol}}\sim |a_1^{(2)}b_1^{(2)}c_1d_1|\cos(\Delta \varphi^{pump})
\Big[\alpha |a_1^{(2)}|^2+\beta |b_1^{(2)}|^2\Big], \label{eq:Direct_nonl_anl}
\end{align}
where we introduced the coefficients
\begin{align}
& \alpha=D_{N_x \rightarrow M_x, M_x}D_{M_y\rightarrow M_x, N_y}
-D_{N_x \rightarrow M_y, M_y}D_{M_y \rightarrow M_y, N_x},\nonumber \\
&\beta=D_{N_x \rightarrow N_x, N_x} D_{M_y \rightarrow N_x, M_y}
-D_{N_x \rightarrow N_y, N_y}D_{M_y \rightarrow N_y, M_x}.\nonumber 
\end{align}

The expression \eqref{eq:Direct_nonl_anl} provides a simple yet clear result on the origin of the two-photon generation directivity. First of all, one need to mention  the factor containing  the phase difference between ED and MD modes at the fundamental wavelength $\Delta\varphi^{pump}= \varphi_{c_1(MD)}-\varphi_{d_1(ED)}$. Thus, the directionality of the collinear two-photon detection can be expected for the same in-phase or out-of-phase conditions as classical  Kerker effect \eqref{eq:Direct_nonl}. Secondly, there is an additional factor in square brackets, which strongly differs two-photon generation from the elastic scattering case. It contains mode amplitudes $a_1$ and $b_1$ multiplied by $\alpha$ and $\beta$ coefficients defined by the amplitudes of the decay. For the cases shown in Fig.~\ref{fig:3_direct_pattern} the factor in the square brackets in \eqref{eq:Direct_nonl_anl} is strictly positive, however it strongly depends on the particular modes involved in the decay process and can freely change its sign. The spectral dependence of $\alpha$- and $\beta$-factors for the parameters used in Fig.\ref{fig:3_direct_pattern} is provided in Fig.\ref{fig:D_coeff_spectra} in Appendix \ref{app:Derivation_W}.    

The directional generation of entangled photons is illustrated in Fig.~\ref{fig:3_direct_pattern}, where in panel (a) the directivity of collinear generation is shown for different pumping wavelengths. The spectrum of forward-to-backward ratio $w^{\t{unpol}}_{\theta=0}/w^{\t{unpol}}_{\theta=\pi}$ 
is shown in Fig.~\ref{fig:3_direct_pattern}~(b) demonstrating strong backward two-photon scattering at $\lambda_p=720$ nm and forward scattering at $\lambda_p=1086$ nm. The corresponding amplitudes and phases of  Mie  coefficients governing the nonlinear decay are depicted  in  panels Fig.~\ref{fig:3_direct_pattern} (b) and (c), which show that in the course of resonant collinear decay the Kerker-type conditions are applied for the pumping dipole modes providing  directional photon generation.

\subsection{Various crystalline structures for directional decay }
The directivity of the photon generation supposed in this paper is based on electric and magnetic dipoles interference. Thus, these modes should be allowed in decaying channels, which in their turn are defined by the crystalline structure of the material.{ There are two rules for directional emission: i) should expect to observe cross-polarized dipoles generation and co-directional dipoles generation in the direction of one of the crossed, $\rightarrow \ve W_{\alpha}, \ve W_{\beta}$ and $\rightarrow \ve W_{\alpha}, \ve W_{\alpha}$, $\alpha \neq \beta$; ii) in the two decays described in point (i) there must be an odd number of electrical dipoles, 1 or 3. It turns out, that depending on the crystalline structure of the material one expect observing directivity of photon emission in arbitrary directions.  Fig.~\ref{fig:Table_chi2} summarizes different crystalline structures of nonlinear materials which  provide directional collinear generation of entangled photons and corresponding dipole channels of decay. One can see that depending on the orientation of crossed dipoles the directivity of the photon emission can be preferrable in $x$-, $y$-, or $z$-axis.}

\subsection{Polarization entangelement}
\label{sec:Polarization}
The photons generated in course of SPDC process demonstrate not only  spatial correlations discussed above, but also polarization correlations which were excluded  from consideration in unpolarized detection rate $w^{unpol}$. In order to analyze the polarization state of generated photons, we will plot the detection probability $w_{\ve d_i, \ve d_s}$ for different orientation of the detectors $\ve d_i$ and $\ve d_s$. The correlated photons generated in configuration shown at Fig.~\ref{fig:Pol_corr} are polarized in the $x-y$ plane, thus we will trace the correlation between $x$ and $y$ polarizations and also between left and right circular polarizations. That can be done by choosing the detecting dipoles $\ve d_{i,s}= d_0 \ve e_{x,y}$ for linear polarizations and $\ve d_{i,s}= d_0/\sqrt{2} (\ve e_{x}\pm i\ve e_{y})$. The detections probability is integrate over finite  solid angle $\theta\in[0^{\circ},60^{\circ}]$ which corresponds to experimental detection scheme. The results of the simulations are shown in Fig.~\ref{fig:Pol_corr} for linear (b) and circular polarizations (c). One can see that the detected photons have dominantly the same $x$-polarizations, which corresponds to the D-coefficients shown in the Fig.~\ref{fig:2_GaAs}(b), where it can be seen that the decays into two electric dipoles directed along x-axis and to the electric dipole along x-axis and magnetic dipole along y-axis are several times higher than all other possible decays. 

\section{Conclusion} 

In this paper, we have proposed a theoretical approach describing the generation of correlated photons through spontaneous down-conversion process in sub-wavelength dielectric resonator supporting low order Mie resonant modes. Using the two-photon amplitude approach, we have identified the mechanism of spontaneous photon decay in terms of electromagnetic multipoles. As a result, we have shown that by proper designing the modes for a particular class of crystalline materials, one can achieve strongly directional emission of correlated photons. For the collinear geometry provided by idler and signal detectors positioned in the same place, we have formulated the conditions of strongly forward/backward photons generation, which surprisingly appeared to be very similar to classical Kerker-effect conditions. 

Finally, the results reported in our work will provide new grounds for quantum state generation via SPDC process in Mie resonant structures relying on the advances of multipole photonics in designing and controlling light emission from subwavelength resonators.

\section*{Acknowledgments}
The work was supported by Russian Foundation for Basic Research, projects numbers 18-02-01206, 20-32-90238. A.S.S. acknowledges support by the Australian Research Council, project number DE180100070. A.A.N. acknowledges support by the Quantum Technology Centre, Faculty of Physics, Lomonosov Moscow State University.  The authors thank Alexander Poddubny and Andrey Sukhorukov for fruitful discussions. 
\appendix
\section{Vector spherical harmonics}
\label{app:VSH}
Vector spherical harmonics used above are defined in \cite{Bohren1998} as $\vec{M}_{^{e}_{o}mn}=\nabla \times (\ve r \psi_{^{e}_{o}mn})$ and $\vec{N}_{^{e}_{o}mn}=\frac{\nabla \times \vec{M}_{^{e}_{o}mn}}{k}$, where\\ $\psi_{emn}=\cos m\varphi P^m_n(\cos \theta) z_n(\rho)$ and $\psi_{omn}=\sin m\varphi P^m_n(\cos \theta) z_n(\rho)$ are the scalar spherical functions, proportional to the tesseral (real) spherical functions.
Magnetic vector spherical harmonics:
\begin{equation}\label{Me}
\begin{aligned}
\ve{M}_{emn}(k, \ve{r})=-\dfrac{m}{\sin\theta}\sin m\varphi\cdot P_n^m(\cos\theta)z_n(\rho)\ve{e}_{\theta}-\\
-\cos m\varphi\cdot \dfrac{dP_n^m(\cos\theta)}{d\theta}z_n(\rho)\ve{e}_{\varphi}
\end{aligned}
\end{equation}

\begin{equation}\label{Mo}
\begin{aligned}
\ve{M}_{omn}(k, \ve{r})=\dfrac{m}{\sin\theta}\cos m\varphi\cdot P_n^m(\cos\theta)z_n(\rho)\ve{e}_{\theta}-\\-\sin m\varphi\cdot \dfrac{dP_n^m(\cos\theta)}{d\theta}z_n(\rho)\ve{e}_{\varphi}
\end{aligned}
\end{equation}
Electric vector spherical harmonics:
\begin{equation}\label{Ne}
\begin{aligned}
\ve{N}_{emn}(k, \ve{r})=\dfrac{z_n(\rho)}{\rho}\cos m\varphi\cdot n(n+1) P_n^m(\cos\theta)\bm{e}_{r}+\\+\cos m\varphi\cdot \dfrac{dP_n^m(\cos\theta)}{d\theta}\dfrac{1}{\rho}\dfrac{d}{d\rho}[\rho z_n(\rho)]\ve{e}_{\theta}- \\
 -m\sin m\varphi\cdot \dfrac{P_n^m(\cos\theta)}{\sin\theta}\dfrac{1}{\rho}\dfrac{d}{d\rho}[\rho z_n(\rho)]\ve{e}_{\varphi}
\end{aligned}
\end{equation}

\begin{equation}
\begin{aligned}\label{No}
\ve{N}_{omn}(k, \ve{r})=\dfrac{z_n(\rho)}{\rho}\sin m\varphi\cdot n(n+1) P_n^m(\cos\theta)\bm{e}_{r}+\\+\sin m\varphi\cdot\dfrac{dP_n^m(\cos\theta)}{d\theta}\dfrac{1}{\rho}\dfrac{d}{d\rho}[\rho z_n(\rho)]\ve{e}_{\theta}+\\+m\cos m\varphi\cdot \dfrac{P_n^m(\cos\theta)}{\sin\theta}\dfrac{1}{\rho}\dfrac{d}{d\rho}[\rho z_n(\rho)]\ve{e}_{\varphi},
\end{aligned}
\end{equation}
where $n=1, 2, 3\dots$, $m=0,\dots,n$, the indices $e$ and $o$ denote even and odd  $\psi$-functions parity with respect to  $\varphi\leftrightarrow -\varphi$ transformation. In place of $z_n(\rho)$, where $\rho=kr$ is the dimensionless variable, can be spherical bessel $j_n(\rho)$ or spherical hankel $h_n^{(1)}(\rho)$ functions, depending on the specific boundary conditions; $P^m_n$ are the associated Legendre polynomials. The coefficients in expressions for the scattered field (\ref{eq:E_s}) and  the field inside nanoparticle (\ref{eq:E_p}) are determined by the following expressions
\begin{gather}
a_n(\omega)=\dfrac{\dfrac{\mu_1}{\mu_2}\Big(\dfrac{k_2}{k_1} \Big)^2j_n(\rho_2)[\rho_1j_n(\rho_1)]'-j_n(\rho_1)[\rho_2j_n(\rho_2)]'}{\dfrac{\mu_1}{\mu_2}\Big(\dfrac{k_2}{k_1} \Big)^2j_n(\rho_2)[\rho_1h_n^{(1)}(\rho_1)]'-h_n^{(1)}(\rho_1)[\rho_2j_n(\rho_2)]'},
\end{gather}
\begin{gather}
b_n(\omega)=\dfrac{ j_n(\rho_2)[\rho_1j_n(\rho_1)]'-\dfrac{\mu_1}{\mu_2}j_n(\rho_1)[\rho_2j_n(\rho_2)]'}{j_n(\rho_2)[\rho_1h_n^{(1)}(\rho_1)]'-\dfrac{\mu_1}{\mu_2}h_n^{(1)}(\rho_1)[\rho_2j_n(\rho_2)]'},
\end{gather}
\begin{gather}
c_n(\omega)=\dfrac{ j_n(\rho_1)[\rho_1h_n^{(1)}(\rho_1)]'-h_n^{(1)}(\rho_1)[\rho_1j_n(\rho_1)]'}{j_n(\rho_2)[\rho_1h_n^{(1)}(\rho_1)]'-\dfrac{\mu_1}{\mu_2}h_n^{(1)}(\rho_1)[\rho_2j_n(\rho_2)]'}
\end{gather}
\begin{gather}
d_n(\omega)=\dfrac{ j_n(\rho_1)[\rho_1h_n^{(1)}(\rho_1)]'-h_n^{(1)}(\rho_1)[\rho_1j_n(\rho_1)]'}{\dfrac{\mu_1}{\mu_2}\dfrac{k_2}{k_1}j_n(\rho_2)[\rho_1h_n^{(1)}(\rho_1)]'-\dfrac{k_1}{k_2} h_n^{(1)}(\rho_1)[\rho_2j_n(\rho_2)]'},
\end{gather}
where $k_1(\omega)={\omega\sqrt{\varepsilon_1}}/{c}$ is the wavenumber outside the nanoparticle,~$k_2(\omega)={\omega\sqrt{\varepsilon_2(\omega)}}/{c}$ is the wavenumber inside the nanoparticle,~$\mu_1=\mu_2=1$ are the magnetic permeabilities of media and nanoparticle,~$\rho_1=k_1a,~\rho_2=k_2a$, where $a$ is the radius of the sphere.\\

To obtain linear (\ref{eq:Linear_Kerker}) and nonlinear (\ref{eq:Direct_nonl_anl}) Kerker conditions, we used  asymptotics of the spherical Hankel functions in the far field $h^{(1)}_n(\rho)\sim\dfrac{(-i)^ne^{i\rho}}{i\rho}$ and $\dfrac{1}{\rho}\dfrac{d}{d\rho}[\rho h^{(1)}_n(\rho)]\sim\dfrac{(-i)^ne^{i\rho}}{\rho}$. Below we show an example of how we calculated the phase difference between the harmonics basing on the  analysis of the scalar product:\\
\begin{equation}
\begin{aligned}
\vec{M}^{(1)}_{y}(\varphi,\theta,\rho)\cdot \vec{N}^{(1)*}_{x}(\varphi,\theta,\rho)=\\
\dfrac{P^1_1(\cos\theta)}{\sin\theta} h_{n}^{(1)}(\rho)\cdot\Big(\dfrac{d P^1_1(\cos \theta)}{d\theta} \dfrac{1}{\rho}\dfrac{d}{d\rho}[\rho h_{n}^{(1)}(\rho)]\Big)^* 
\sim\\ \sim\cos\theta\dfrac{(-i)^n e^{i\rho}}{i\rho}\cdot \dfrac{(i)^n e^{-i\rho}}{\rho}
\sim\cos\theta\dfrac{-i}{\rho^2}.
\end{aligned}
\end{equation}

Here, we used that $P_1^1(\cos\theta)=\sin \theta$. Thus, this scalar product has equal value and different signs at $\theta=0$  and at $\theta=\pi$. Aslo note, that it is proportional to the angular part of $\psi_{e01}$ and the phase differences are  $\varphi_{\ve M_y}-\varphi_{\ve N_x}=-\dfrac{\pi}{2}$ at $\theta=0$ and $\varphi_{\ve M_y}-\varphi_{\ve N_x}=\dfrac{\pi}{2}$ at $\theta=\pi$.\\

To obtain the required dyadic Green's function for the considered system, the following set of equations was solved \cite{Mason}:
\begin{widetext}
\begin{equation}
\left\{\begin{aligned}
&\bigtriangledown\times\bigtriangledown\times\vec{\hat{G}}(\vec{r},\vec{r'},\omega)-k_2^2\vec{\hat{G}}(\vec{r},\vec{r'},\omega)=4\pi\Big(\dfrac{\omega}{c}\Big)^2
\hat{1}\delta(\vec{r}-\vec{r'}), ~r\leq a\\&
\bigtriangledown\times\bigtriangledown\times\vec{\hat{G}}(\vec{r},\vec{r'},\omega)-k_1^2\vec{\hat{G}}(\vec{r},\vec{r'},\omega)=0, ~r\geq a
\end{aligned}\right.
\end{equation}

And the answer for the dyadic Green function pertaining for a source in the presence of a dielectric sphere, the source is inside the sphere, and the observer is outside

\begin{equation}
\begin{aligned}
\vec{\hat{G}}(\vec{r},\vec{r}',\omega)=4\pi\Big(\dfrac{\omega}{c}\Big)^2\dfrac{ik_2(\omega)}{4\pi}\sum\limits_{n=1}^{\infty}\sum\limits_{m=0}^n(2-\delta_0)\dfrac{2n+1}{n(n+1)}\dfrac{(n-m)!}{(n+m)!}\cdot \\\cdot [a_n^{(2)}\vec{M}^{(1)}_{\substack{e\\o}mn}(k_1(\omega), \vec{r})\otimes\vec{M}_{\substack{e\\o}mn}(k_2(\omega),\vec{r}')+b_n^{(2)}\vec{N}^{(1)}_{\substack{e\\o}mn}(k_1(\omega),\vec{r})\otimes\vec{N}_{\substack{e\\o}mn}(k_2(\omega),\vec{r}')],
\end{aligned}
\label{eq:Green}
\end{equation}
\end{widetext}
where the index ${\rm (1)}$ in harmonics means that the spherical Hankel function of first kind is used, $\delta_0=1$ when $m=0$, $\delta_0=0$ when $m\neq 0$, and the coefficients are determined by the following expressions:
\begin{equation}
a^{(2)}_n(\omega)=\dfrac{ h_n^{(1)}(\rho_2)[\rho_2j_n(\rho_2)]'-j_n(\rho_2)[\rho_2 h_n^{(1)}(\rho_2)]'}{h_n^{(1)}(\rho_1)[\rho_2j_n(\rho_2)]'-\dfrac{\mu_2}{\mu_1} j_n(\rho_2)[\rho_1 h_n^{(1)}(\rho_1)]'}
\end{equation}
\begin{equation}
b^{(2)}_n(\omega)=\dfrac{ h_n^{(1)}(\rho_2)[\rho_2j_n(\rho_2)]'-j_n(\rho_2)[\rho_2 h_n^{(1)}(\rho_2)]'}{\dfrac{\mu_2}{\mu_1}\dfrac{k_1}{k_2} h_n^{(1)}(\rho_1)[\rho_2j_n(\rho_2)]'-\dfrac{k_2}{k_1} j_n(\rho_2)[\rho_1 h_n^{(1)}(\rho_1)]'}
\end{equation}

\section{Derivation of two-photon amplitude}
\label{app:Derivation_T}
To get an expression for the two-photon amplitude in terms of VSH (\ref{eq:T_harmonics}) we substitute dyadic Green functions from previous section (\ref{eq:Green}) and $\ve E_p$ pump field inside the nanoparticle (\ref{eq:E_p}) to the expression (\ref{eq:T_general}), where we  apply the new notation for the VSH.
Dyadic Green function for the idler photon coordinate as observer
$\vec{\hat{G}}(\vec{r}_i,\vec{r}_0,\omega_i)=ik_{i2}\Big(\dfrac{\omega_i}{c}\Big)^2\sum\limits_{\ve J_i}A_{\ve J_i} \ve W_{\ve J_i}^{(1)}(k_{i1}, \ve r_i)\otimes \ve W_{\ve J_i}(k_{i2}, \ve r_0)$
and for the signal photon coordinate as observer
$\vec{\hat{G}}(\vec{r}_0,\vec{r}_s,\omega_s)=ik_{s2}\Big(\dfrac{\omega_s}{c}\Big)^2\sum\limits_{\ve J_s}A_{\ve J_s} \ve W_{\ve J_s}(k_{s2}, \ve r_0)\otimes \ve W_{\ve J_s}^{(1)}(k_{s1}, \ve r_s)$, the nonlinear generation matrix is expressed in terms of VSH as follows:
$\hat{\Gamma}=\hat\chi_{} \vec E_p^{}(\ve r_0)=\sum\limits_{\ve J_p} B_{\ve J_p}\hat\chi_{} \ve W_{\ve J_p}(k_2, \ve r_0),$
where \\$A_{\ve J}=(2-\delta_0)\dfrac{2n+1}{n(n+1)}\dfrac{(n-m)!}{(n+m)!}\cdot\begin{cases}
a_n^{(2)}\text{,~if}~p_i\rightarrow M \\b_n^{(2)}\text{,~if}~p_i\rightarrow N
\end{cases}$ \\and $B_{\ve J	}=E_n\cdot\begin{cases}
c_n\text{,~if}~p_i\rightarrow M\\-id_n\text{,~if}~p_i\rightarrow N\end{cases}.$\\
Performing the substitution, we obtain the two-photon amplitude represented through VSH:
\begin{widetext}
\begin{equation}
    \begin{aligned}
T_{is}(\vec{r}_i,\omega_i,\vec{d}_i;\vec{r}_s,\omega_s,\vec{d}_s)=\int\limits_V \ve d_i^*\vec{\hat{G}}(\vec{r}_i,\vec{r}_0,\omega_i)\hat{\Gamma}(\ve r_0) \vec{\hat{G}}(\vec{r}_0,\vec{r}_s,\omega_i)  \ve d_s^*d^3 r_0 =\sum\limits_{\ve J_p, \ve J_i, \ve J_s} \underbrace{ik_{i2}\Big(\dfrac{\omega_i}{c}\Big)^2A_{\ve J_i}ik_{s2}\Big(\dfrac{\omega_s}{c}\Big)^2A_{\ve J_s}B_{\ve J_p}}_{\widetilde T_{\ve J_p \rightarrow\ve J_i, \ve J_s}}\times
\\\times\int\limits_V \ve W_{\ve J_i}(k_{i2}, \ve r_0) \hat\chi_{}\vec W_{\ve J_p}(k_2,\ve r_0)\ve W_{\ve J_s}(k_{s2}, \ve r_0) d^3r_0 \times \Big(\ve d_i^*\cdot \ve W_{\ve J_i}^{(1)}(k_{i1}, \ve r_i)\Big)\cdot\Big(\ve W_{\ve J_s}^{(1)}(k_{s1}, \ve r_s) \cdot \ve d_s^*\Big)=
\\=\sum\limits_{\ve J_p, \ve J_i, \ve J_s} \widetilde T_{\ve J_p \rightarrow\ve J_i, \ve J_s}\times  
\\
\times\underbrace{\int\limits_V  W_{\ve J_i,\alpha}(k_{i2}, \ve r_0) \chi_{\alpha\beta\gamma}W_{\ve J_p, \gamma}(k_2,\ve r_0) W_{\ve J_s,\beta}(k_{s2}, \ve r_0) d^3r_0}_{D_{\ve J_p\rightarrow \ve J_i, \ve J_s}} \times \Big(\ve d_i^*\cdot \ve W_{\ve J_i}^{(1)}(k_{i1}, \ve r_i)\Big)\cdot\Big(\ve W_{\ve J_s}^{(1)}(k_{s1}, \ve r_s) \cdot \ve d_s^*\Big)= 
\\=\sum\limits_{\ve J_p, \ve J_i, \ve J_s} \widetilde T_{\ve J_p \rightarrow\ve J_i, \ve J_s}\times D_{\ve J_p\rightarrow \ve J_i, \ve J_s}\times \Big(\ve d_i^*\cdot \ve W_{\ve J_i}^{(1)}(k_{i1}, \ve r_i)\Big)\cdot\Big(\ve W_{\ve J_s}^{(1)}(k_{s1}, \ve r_s) \cdot \ve d_s^*\Big),
 \end{aligned}
 \end{equation}
\end{widetext}
where $W_{\ve J,\alpha}(k, \ve r)$ is the projection of the vector $\ve W_{\ve J}(k, \ve r)$ on the $\alpha$-axis, $\alpha=x,y,z$. We also use the following notation $\widetilde T_{\ve J_p \rightarrow\ve J_i, \ve J_s}=-\Big(\dfrac{\omega_i\omega_s}{c^2}\Big)^2k_{i2}k_{s2}A_{\ve J_i}A_{\ve J_s}B_{\ve J_p}$ and $D_{\ve J_p\rightarrow \ve J_i, \ve J_s}=\int\limits_V  W_{\ve J_i,\alpha}(k_{i2}, \ve r_0) \chi_{\alpha\beta\gamma}W_{\ve J_p, \gamma}(k_2,\ve r_0) W_{\ve J_s,\beta}(k_{s2}, \ve r_0) d^3r_0$.\\

\section{Derivation of unpolarized counting rate}
\label{app:Derivation_W}
Let us now provide the derivation of the expression (\ref{eq:Direct_nonl_anl}). First, we’ll show how we summed up over $\ve d_i$ and $\ve d_s$ in the expression for counting rate (\ref{eq:Direct_nonl}):
\begin{widetext}
\begin{equation}
    \begin{aligned}
w_{is}^{{unpol}}(\ve{r}_i,\omega_i;\ve{r}_s,\omega_s)= \dfrac{2\pi}{\hbar}\sum\limits_{\ve d_i, \ve d_s} \left|T_{is}(\ve{r}_i,\omega_i,\ve{d}_i;\ve{r}_s,\omega_s,\ve{d}_s)\right|^2=\\
=\dfrac{2\pi}{\hbar}\sum\limits_{\ve d_i, \ve d_s}
 \sum\limits_{\substack{\ve J_p, \ve J_i, \ve J_s\\ \ve J'_p, \ve J'_i, \ve J'_s }}  \underbrace{\widetilde T_{\ve J_p \rightarrow \ve J_i, \ve J_s}\cdot D_{\ve J_p \rightarrow \ve J_i, \ve J_s}\cdot\widetilde T^*_{\ve J'_p \rightarrow \ve J'_i, \ve J'_s}\cdot  D^*_{\ve J'_p \rightarrow \ve J'_i, \ve J'_s}}_{C^{\ve J_p \rightarrow \ve J_i, \ve J_s}_{ \ve J'_p \rightarrow \ve J'_i, \ve J'_s}}\times\\\times 
 \left(\ve d_i^*\cdot \ve W^{(1)}_{\ve J_i}(k_{i1},\ve{r_i})\right)\times\left(\ve W^{(1)}_{\ve J_s}(k_{s1},\ve{r_s})\cdot \ve d_s^*\right)\times
\left(\ve d_i\cdot \ve W^{*(1)}_{\ve J_i'}(k_{i1},\ve{r_i})\right)\times\left(\ve W^{*(1)}_{\ve J_s'}(k_{s1},\ve{r_s})\cdot \ve d_s\right).
     \end{aligned}
\end{equation}

 Since we sum over all possible polarizations $\ve d_{i(s)}= d_{i(s)}\ve e_{\alpha}$, where $\alpha=x,y,z$, we can assume that the scalar product $\ve d^*_{i(s)}\cdot \ve W^{(1)}_{\ve J_{i(s)}}$ is the projection of the vector $\ve W^{(1)}_{\ve J_{i(s)}}$ on the $\alpha$-axis $W^{(1)}_{\ve J_{i(s)}, \alpha}$. In this way, $\sum\limits_{\ve d_{i(s)}}(\ve d_{i(s)}^*\cdot \ve W^{(1)}_{\ve J_{i(s)}})(\ve d_{i(s)}\cdot \ve W^{(1)*}_{\ve J'_{i(s)}})=\sum\limits_{\alpha}W^{(1)}_{\ve J_{i(s)}, \alpha}W^{(1)*}_{\ve J'_{i(s)}, \alpha}=\ve W^{(1)}_{\ve J_{i(s)}}\cdot \ve W^{(1)*}_{\ve J'_{i(s)}}.$

 Hence,
 \begin{equation}
w_{is}^{{unpol}}(\ve{r}_i,\omega_i;\ve{r}_s,\omega_s)=\dfrac{2\pi}{\hbar} \sum\limits_{\substack{\ve J_p, \ve J_i, \ve J_s\\\ve J'_p, \ve J'_i, \ve J'_s}} C^{\ve J_p \rightarrow \ve J_i, \ve J_s}_{ \ve J'_p \rightarrow \ve J'_i, \ve J'_s} \left( \ve W^{(1)}_{\ve J_i}(k_{i1},\ve{r_i})\cdot \ve W^{*(1)}_{\ve J_i'}(k_{i1},\ve{r_i}) \right)
\times\left(\ve W^{(1)}_{\ve J_s}(k_{s1},\ve{r_s})\cdot \ve W^{*(1)}_{\ve J_s'}(k_{s1},\ve{r_s}) \right),
 \label{eq:W_unpol_app}
     \end{equation}
 where $C^{\ve J_p \rightarrow \ve J_i, \ve J_s}_{ \ve J'_p \rightarrow \ve J'_i, \ve J'_s}=\widetilde T_{\ve J_p \rightarrow \ve J_i, \ve J_s}\cdot D_{\ve J_p \rightarrow \ve J_i, \ve J_s}\cdot\widetilde T^*_{\ve J'_p \rightarrow \ve J'_i, \ve J'_s}\cdot  D^*_{\ve J'_p \rightarrow \ve J'_i, \ve J'_s}$.\\
  \end{widetext}
 Next, to obtain eq. (\ref{eq:Direct_nonl_anl}), 
    \begin{figure}[h!]
\center{\includegraphics[width=1\linewidth]{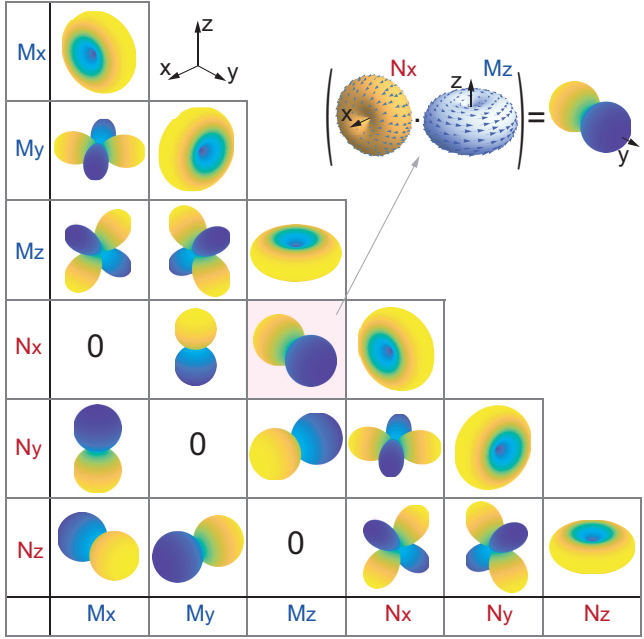}}
\caption{All possible scalar products $|\ve W_{\ve J} \cdot \ve W_{\ve J'}|$ of dipole vector spherical harmonics.} 
\label{fig:app_scalar}
\end{figure} 
we have considered all the dipole terms in (\ref{eq:W_unpol_app}) and found $w_{is}^{{unpol}}(\ve{r},\omega_p/2)$ at $\theta=0,\pi$. Since the expression (\ref{eq:W_unpol_app}) includes scalar products of VSH (all the possible options are shown in Figure \ref{fig:app_scalar}), it can be seen that the following scalar products are non-zero in the $\theta = 0, \pi$ directions: $\ve M_x \cdot \ve N_y$, $\ve M_y \cdot \ve N_x, |\ve N_x|^2, |\ve N_y|^2, |\ve M_x|^2, |\ve M_y|^2$. For directivity along the $z$-axis, one of two scalar products must enter with the same sign in the $\theta=0$ and $\theta=\pi$ directions, i.e. $|\ve N_x|^2, |\ve N_y|^2, |\ve M_x|^2, |\ve M_y|^2$. And the second scalar product should have different signs in these two directions as $\ve M_x \cdot \ve N_y$ or $\ve M_y \cdot \ve N_x$. Note, that each product contains only just idler or signal photons. {So the possible decays are $\ve M_x\ve N_x+\ve N_y\ve N_x$, $\ve M_x \ve N_y+\ve N_y \ve N_y$, $\ve M_x \ve M_y+\ve N_y \ve M_y$, $\ve M_x \ve M_x+\ve N_y \ve M_x$; $\ve M_y \ve N_x+\ve N_x \ve N_x$, $\ve M_y \ve N_y+\ve N_x \ve N_y$, $\ve M_y \ve M_y+\ve N_x \ve M_y$, $\ve M_y \ve M_x+\ve N_x\ve M_x$. 
\begin{figure}[t]
\center{\includegraphics[width=1.5\linewidth]{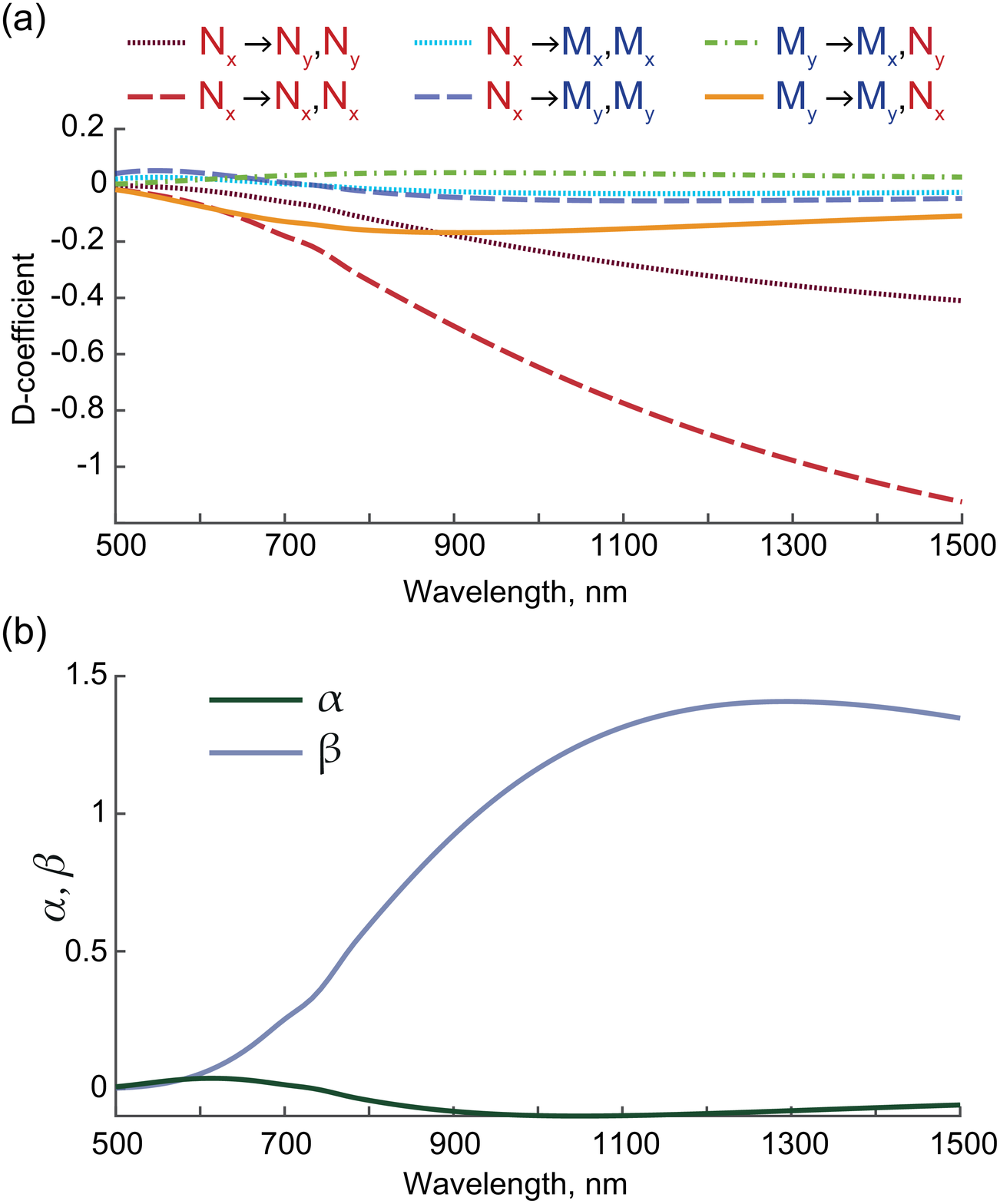}}
\caption{(a) Dependence of overlapping coefficients $D_{\ve J_p \rightarrow \ve J_i, \ve  J_s}$ on the fundamental wavelength $\lambda_p$. Dotted light blue -- decay into two magnetic dipoles along the x-axis, dashed blue -- decay into two magnetic dipoles along the y-axis, dashed red -- decay into two electric dipoles along the x-axis, dotted purple -- decay into two electric dipoles along the y-axis, dashed-dot green -- decay into crossed electric dipole along y axis and magnetic dipole along the x-axis, solid yellow -- decay into crossed electric dipole along the x-axis and magnetic dipole along the y-axis.(b) Dependence  $\alpha=D_{\ve M_y \rightarrow \ve M_x, \ve  N_y}D_{ \ve N_x \rightarrow \ve  M_x, \ve M_x}-D_{\ve M_y \rightarrow \ve M_y, \ve  N_x}D_{ \ve N_x \rightarrow \ve  M_y, \ve M_y}$ (solid blue) and $\beta=D_{\ve M_y \rightarrow \ve M_y, \ve  N_x}D_{ \ve N_x \rightarrow \ve  N_x, \ve N_x}-D_{\ve M_y \rightarrow \ve M_x, \ve  N_y}D_{ \ve N_x \rightarrow \ve  N_y, \ve N_y}$   (solid red) coefficients on the fundamental wavelength $\lambda_p$ }
\label{fig:D_coeff_spectra}
\end{figure}
From this, it follows that for the directional emission along the $z$-axis, we need two types of decay simultaneously: one is both photons into two crossed dipoles (along the $x$- and $y$- axes) and the second into two co-directional (both along the $x$- or $y$- axes). And there must also be an odd number of electrical dipoles: 1 or 3. Let us now consider the geometry, depicted in Fig. \ref{fig:2_GaAs}. Here the optical axis of the wurzite crystalline structure is oriented along the x-axis, so the tensor has only "green" components. So only part of transitions we are interested in is allowed by the selection rules.}  Let us substitute them in the expression \ref{eq:W_unpol_app}. For the collinear decay we get $w_{is}^{{unpol}}(\ve{r},\omega_p/2)$ and it can be represented by two terms: phase-independent part $w_{0}$ and phase-dependent $w_{is}^{{unpol}}(\ve{r},\omega_p/2)-w_{0}$:\\
 \begin{equation}
 \begin{aligned}
& w_{0}=\dfrac{2\pi}{\hbar}\left(2C^{\ve M_y \rightarrow \ve M_y, \ve N_x}_{ \ve M_y \rightarrow \ve M_y, \ve N_x}  |\ve M^{(1)}_{y}|^2\cdot |\ve N^{(1)}_{x}|^2+\right.\\
&+2C^{\ve M_y \rightarrow \ve N_x, \ve M_y}_{ \ve M_y \rightarrow \ve M_y, \ve N_x} | \ve N^{(1)}_{x}\cdot \ve M^{*(1)}_{y}|^2 +\\
&+2C^{\ve M_y \rightarrow \ve M_x, \ve N_y}_{ \ve M_y \rightarrow \ve M_x, \ve N_y}  |\ve M^{(1)}_{x}|^2|\ve N^{(1)}_{y}|^2+\\
&+2C^{\ve M_y \rightarrow \ve N_y, \ve M_x}_{ \ve M_y \rightarrow \ve M_x, \ve N_y} |\ve N^{(1)}_{y}\cdot \ve M^{*(1)}_{x}|^2+
+C^{\ve N_x \rightarrow \ve N_x, \ve N_x}_{ \ve N_x \rightarrow \ve N_x, \ve N_x} |\ve N^{(1)}_{x}|^4+\\
&+C^{\ve N_x \rightarrow \ve N_y, \ve  N_y}_{ \ve N_x \rightarrow \ve  N_y, \ve N_y} | \ve N^{(1)}_{y}|^4+C^{\ve N_x \rightarrow \ve M_x, \ve M_x}_{ \ve N_x \rightarrow \ve M_x, \ve M_x} |\ve M^{(1)}_{x}|^4+\\
&+C^{\ve N_x \rightarrow \ve M_y, \ve  M_y}_{ \ve N_x \rightarrow \ve  M_y, \ve M_y} | \ve M^{(1)}_{y}|^4+(C^{\ve N_x \rightarrow \ve N_x, \ve  N_x}_{ \ve N_x \rightarrow \ve  M_y, \ve M_y} |\ve N^{(1)}_{x}\cdot \ve M^{*(1)}_{y}|^2+\\
&\left.+C^{\ve N_x \rightarrow \ve N_y, \ve  N_y}_{ \ve N_x \rightarrow \ve  M_x, \ve M_x} |\ve N^{(1)}_{y}\cdot \ve M^{*(1)}_{x}|^2+c.c.\right).
 \end{aligned}
 \end{equation}
 Next, we write down the part depending on the direction:
 \begin{equation}
 \begin{aligned}
&w_{is}^{{unpol}}(\ve{r},\omega_p/2)-w_{0}= \dfrac{2\pi}{\hbar}4\Re[C^{\ve M_y \rightarrow \ve M_y, \ve  N_x}_{ \ve N_x \rightarrow \ve  N_x, \ve N_x} \left( \ve M^{(1)}_{y}\cdot \ve N^{*(1)}_{x}\right)|N^{(1)}_{x}|^2\\
&+C^{\ve M_y \rightarrow \ve M_y, \ve  N_x}_{ \ve N_x \rightarrow \ve  M_y, \ve M_y}\left(\ve N^{(1)}_{x}\cdot \ve M^{*(1)}_{y}\right)|M^{(1)}_{y}|^2+\\
&+C^{\ve M_y \rightarrow \ve M_x, \ve  N_y}_{ \ve N_x \rightarrow \ve  N_y, \ve N_y} \left( \ve M^{(1)}_{x}\cdot \ve N^{*(1)}_{y}\right)|N^{(1)}_{y}|^2+\\
&+C^{\ve M_y \rightarrow \ve M_x, \ve  N_y}_{ \ve N_x \rightarrow \ve  M_x, \ve M_x}\left(\ve N^{(1)}_{y}\cdot \ve M^{*(1)}_{x}\right)|M^{(1)}_{x}|^2]
 \end{aligned}
 \end{equation}

 Then, we use relation $\Re[a\cdot b]=\Re[a]\Re[b]-\Im[a]\Im[b]$ and $\Re[ab^*] = |ab^*| \cos(\varphi_a-\varphi_b)$, $\Im[ab^*] = |ab^*| \sin(\varphi_a-\varphi_b)$ and $\Re [\ve  M_y^{(1)}\cdot \ve N_x^{*(1)}]=0$ at $\theta=0, \pi$ directions.
 Also, using that the coefficients $C^{\ve J_p \rightarrow \ve J_i, \ve  J_s}_{ \ve J'_p \rightarrow \ve  J'_i, \ve J'_s} $ do not depend on the direction, we write down the difference between up and down direction $\Delta w^{unpol}$, where we have excluded all modules of scalar products of VSH, since they are equal in the directions $\theta=0, \pi$:
 
 \begin{widetext}
 \begin{equation}
 \begin{aligned}
     \Delta w^{unpol}\sim 4\Im[C^{\ve M_y \rightarrow \ve M_y, \ve  N_x}_{ \ve N_x \rightarrow \ve  M_y, \ve M_y}-C^{\ve M_y \rightarrow \ve M_y, \ve  N_x}_{ \ve N_x \rightarrow \ve  N_x, \ve N_x}]
     \cdot(\sin(\varphi_{ \ve M_{y}}-\varphi_{\ve N_{x}})|_{\theta=0}-\sin(\varphi_{ \ve M_{y}}-\varphi_{\ve N_{x}})|_{\theta=\pi})+\\
+4\Im[-C^{\ve M_y \rightarrow \ve M_x, \ve  N_y}_{ \ve N_x \rightarrow \ve  N_y, \ve N_y} +C^{\ve M_y \rightarrow \ve M_x, \ve  N_y}_{ \ve N_x \rightarrow \ve  M_x, \ve M_x}]
\cdot(\sin(\varphi_{\ve M_{x}}-\varphi_{ \ve N_{y}}|_{\theta=0}-\sin(\varphi_{\ve M_{x}}-\varphi_{ \ve N_{y}}|_{\theta=\pi})=\\
=8\Im[-C^{\ve M_y \rightarrow \ve M_x, \ve  N_y}_{ \ve N_x \rightarrow \ve  N_y, \ve N_y} +C^{\ve M_y \rightarrow \ve M_x, \ve  N_y}_{ \ve N_x \rightarrow \ve  M_x, \ve M_x}-C^{\ve M_y \rightarrow \ve M_y, \ve  N_x}_{ \ve N_x \rightarrow \ve  M_y, \ve M_y}+C^{\ve M_y \rightarrow \ve M_y, \ve  N_x}_{ \ve N_x \rightarrow \ve  N_x, \ve N_x}]=\\
=8\varepsilon_2\left(\dfrac{\omega_p}{2c}\right)^6\left(\dfrac{3}{2}\right)^6E_0^2\cdot \Im[(ia_1^{(2)}b_1^{(2)}c_1)D_{\ve M_y \rightarrow \ve M_x, \ve  N_y}(-id_1ib_1^{(2)}b_1^{(2)})^*D_{ \ve N_x \rightarrow \ve  N_y, \ve N_y}-\\-(ia_1^{(2)}b_1^{(2)}c_1)D_{\ve M_y \rightarrow \ve M_x, \ve  N_y}(-id_1ia_1^{(2)}a_1^{(2)})^*D_{ \ve N_x \rightarrow \ve  M_x, \ve M_x}+\\
+(ia_1^{(2)}b_1^{(2)}c_1)D_{\ve M_y \rightarrow \ve M_y, \ve  N_x}(-id_1ia_1^{(2)}a_1^{(2)})^*D_{ \ve N_x \rightarrow \ve  M_y, \ve M_y}-\\
(ia_1^{(2)}b_1^{(2)}c_1)D_{\ve M_y \rightarrow \ve M_y, \ve  N_x}(-id_1ib_1^{(2)}b_1^{(2)})^*D_{ \ve N_x \rightarrow \ve  N_x, \ve N_x}]\\
 \end{aligned}
 \end{equation}

 \begin{equation}
 \begin{aligned}
\Delta w^{unpol}\sim8\varepsilon_2\left(\dfrac{\omega_p}{2c}\right)^6\left(\dfrac{3}{2}\right)^6E_0^2|a_1^{(2)}b_1^{(2)}c_1d_1|\times \\ \times\left[|b_1^{(2)}|^2(D_{\ve M_y \rightarrow \ve M_x, \ve  N_y}D_{ \ve N_x \rightarrow \ve  N_y, \ve N_y}-D_{\ve M_y \rightarrow \ve M_y, \ve  N_x}D_{ \ve N_x \rightarrow \ve  N_x, \ve N_x})\sin(\varphi_{c_1}-\varphi_{d_1}+\pi/2)+\right.\\
\left.+|a_1^{(2)}|^2(D_{\ve M_y \rightarrow \ve M_y, \ve  N_x}D_{ \ve N_x \rightarrow \ve  M_y, \ve M_y}-D_{\ve M_y \rightarrow \ve M_x, \ve  N_y}D_{ \ve N_x \rightarrow \ve  M_x, \ve M_x})\sin(\varphi_{c_1}-\varphi_{d_1}+\pi/2)\right]\sim\\
\sim \left(\dfrac{\omega_p}{2c}\right)^6|a_1^{(2)}b_1^{(2)}c_1d_1|\cdot\left[\alpha|a_1^{(2)}|^2+\beta|b_1^{(2)}|^2\right]\cos(\Delta\varphi^{pump})
 \end{aligned}
  \end{equation}
\end{widetext}
where $\alpha=D_{\ve M_y \rightarrow \ve M_x, \ve  N_y}D_{ \ve N_x \rightarrow \ve  M_x, \ve M_x}-D_{\ve M_y \rightarrow \ve M_y, \ve  N_x}D_{ \ve N_x \rightarrow \ve  M_y, \ve M_y}$ and $\beta=D_{\ve M_y \rightarrow \ve M_y, \ve  N_x}D_{ \ve N_x \rightarrow \ve  N_x, \ve N_x}-D_{\ve M_y \rightarrow \ve M_x, \ve  N_y}D_{ \ve N_x \rightarrow \ve  N_y, \ve N_y}$ as mentioned earlier. The $\alpha$ and $\beta$ coefficients are shown in Figure \ref{fig:D_coeff_spectra} as well as all the overlapping integrals $D_{\ve J_p \rightarrow \ve J_i, \ve  J_s}$ used in them. Note that the $\beta$ coefficient is several orders of magnitude greater than the $\alpha$, which means that the main contribution into directive decay is provided  by the electric modes. 
 \newpage

\bibliography{MieSPDC_bib2}

\end{document}